\documentclass [reprint,aps,amsmath,amssymb,twoside,prd,showkeys,superscriptaddress]{revtex4-1}
\usepackage{verbatim}
\usepackage{graphicx}
\usepackage[T1]{fontenc}
\usepackage{textcomp}
\usepackage{txfonts}
\usepackage{tensor}
\usepackage{cancel}
\usepackage{perpage} 
\MakePerPage{footnote} 
\usepackage{color}
\newcommand{\HCd}{\mathcal{H}}
\newcommand{\HCdtD}{\tilde{\HCd}_{\mathrm{Dyn}}}
\newcommand{\HCdD}{\HCd_{\mathrm{Dyn}}}

\def\HCdtD{\tilde{\HCd}_{\mathrm{Dyn}}}
\def\HCdD{\HCd_{\mathrm{Dyn}}}
\def\HCdt0{\tilde{\HCd}_{0}}

\newcommand{\onehalf}{{\textstyle\frac{1}{2}}}

\newcommand{\pfrac}[2]{\frac{\partial{#1}}{\partial{#2}}}
\newcommand{\ppfrac}[3]{\frac{\partial^{2}{#1}}{\partial{#2}\partial{#3}}}
\newcommand{\afffias}{Frankfurt Institute for Advanced Studies (FIAS), Ruth-Moufang-Strasse~1, 60438 Frankfurt am Main, Germany}
\newcommand{\affjwg}{Fachbereich Physik, Goethe-Universit\"at, Max-von-Laue-Strasse~1, 60438~Frankfurt am Main, Germany}
\newcommand{\affgsi}{GSI Helmholtzzentrum f\"ur Schwerionenforschung GmbH, Planckstrasse~1, 64291 Darmstadt, Germany}
\newcommand{\affbgu}{Physics Department, Ben-Gurion University of the Negev, Beer-Sheva 84105, Israel}
\newcommand{\affbahamas}{Bahamas Advanced Study Institute and Conferences, 4A Ocean Heights, Hill View Circle, Stella Maris, Long Island, The Bahamas}

\begin{document}

\title{Quadratic curvature theories formulated as Covariant Canonical Gauge theories of Gravity}

\author{David Benisty}
\email{benidav@post.bgu.ac.il}
\affiliation{\afffias}\affiliation{\affjwg}\affiliation{\affbgu}
\author{Eduardo I. Guendelman}
\email{guendel@bgu.ac.il}
\affiliation{\afffias}\affiliation{\affbgu}\affiliation{\affbahamas}
\author{David Vasak}\email{vasak@fias.uni-frankfurt.de}
\affiliation{\afffias}\affiliation{\affjwg}
\author{Jurgen Struckmeier}\email{struckmeier@fias.uni-frankfurt.de}
\affiliation{\afffias}\affiliation{\affjwg}\affiliation{\affgsi}
\author{Horst Stoecker}\email{stoecker@fias.uni-frankfurt.de}
\affiliation{\afffias}\affiliation{\affjwg}\affiliation{\affgsi}
\begin{abstract}
The Covariant Canonical Gauge theory of Gravity is generalized by including  at the Lagrangian level all possible quadratic curvature invariants. In this approach, the covariant Hamiltonian principle and the canonical transformation framework are applied to derive a Palatini type gauge theory of gravity. The metric $g_{\mu\nu}$, the affine connection $\gamma\indices{^{\lambda}_{\mu\nu}}$ and their respective conjugate momenta, $k^{\mu\nu\sigma}$ and $q\indices{_{\eta}^{\alpha\xi\beta}}$ tensors, are the independent field components describing the gravity. The metric is the basic dynamical field, and the connection is the gauge field. The torsion-free and metricity-compatible version of the space-time Hamiltonian is built from all possible invariants of the $q\indices{_{\eta}^{\alpha\xi\beta}}$ tensor components up to second order. These correspond in the Lagrangian picture to Riemann tensor invariants of the same order. We show that the quadratic tensor invariant is necessary for  constructing the canonical momentum field from the gauge field derivatives, and hence for transforming between Hamiltonian and Lagrangian pictures. Moreover, the theory is extended by dropping metric compatibility and enforcing conformal invariance. This approach could be used for the quantization of the quadratic curvature theories, as for example in the case of conformal gravity. 
\end{abstract}
\maketitle
\section*{Introduction}
A natural way to achieve inflation is considering higher-order curvature corrections in the Hilbert-Einstein Lagrangian as the $R^2$ Starobinsky model \cite{Starobinsky:1980te}, or  other invariants as $R_{\mu\nu}R^{\mu\nu}$ and  $R^{\alpha\beta\gamma\delta}R_{\alpha\beta\gamma\delta}$ \cite{Maeda:2004vm}-\cite{Myrzakulov:2014hca}. Hamiltonian formulations of metric theories with higher curvature terms are problematic as they lead to fourth order derivatives in the equations of motion. The Covariant Canonical Gauge theory of Gravity (CCGG) can overcome this difficulty \cite{Struckmeier:2017vkf}.

The CCGG framework ensures by construction that the action principle is maintained in its form requiring all transformations of a given system to be canonical. The imposed requirement of invariance of the original action integral with respect to local transformations in curved space-time is achieved by introducing additional degrees of freedom, the gauge fields. At the basis of the formulation are two independent fields: the metric $g^{\alpha\beta}$, which encodes the information about lengths and angles of space-time, and the affine connection $\gamma\indices{^{\lambda}_{\alpha\beta}}$, encoding how a vector transforms under parallel displacement. In this formulation, referred to as the Affine-Palatini formalism (or the first-order formalism), these two fields are assumed to be independent dynamical quantities in the action. In addition to those fields there are two "momentum fields": $\tilde{k}^{\alpha\beta\mu}$ which is the conjugate momentum of the metric $g_{\alpha\beta}$, and $\tilde{q}\indices{_{\lambda}^{\alpha\beta\mu}}$, the conjugate momentum of the affine connection $\gamma\indices{^{\lambda}_{\alpha\beta}}$. In the second-order formalism, the connection is assumed to be the Levi-Civita or Christoffel symbol
\begin{equation}
\gamma\indices{^{\rho}_{\mu\nu}} = \left\{ \genfrac{}{}{0pt}{}{\rho}{\mu \nu} \right\} = \onehalf g^{\rho\lambda} (g_{\lambda\mu,\nu}+g_{\lambda\nu,\mu}-g_{\mu\nu,\lambda}),
\end{equation}
leaving the metric as the only a priori independent field. In the Einstein-Hilbert action, where torsion of space-time is neglected anyway, both formulations yield the same equations of motion, and the connection will be in both cases the Christoffel symbol. 

Gauge theories of gravity exploring the covariance of the action with respect to the Lorentz, Poincare and diffeomorphism groups have been considered earlier, see Refs. \cite{Kibble:1961ba}-\cite{Hayashi:1981mm}.  However,  the novel feature facilitated by the covariant Hamiltonian canonical transformation theory \cite{Struckmeier:2016kzt} is the unambiguous derivation from first principles of the coupling of matter fields with dynamical space-time. (The results of the CCGG framework were partially anticipated in Ref. \cite{nester}, though.)

Recently the torsion-free version of CCGG was proven to exploit the correspondence between the first and the second-order formulation by imposing metricity as a constraint implemented via a Lagrange multiplier \cite{Benisty:2018fgu}:  
\begin{equation}\label{t}
\mathcal{L}(g,\gamma) + k^{\alpha\beta\lambda} g_{\alpha\beta;\lambda} \,_{1^{st}\textbf{order}} \Leftrightarrow \mathcal{L}(g)\,_{2^{nd}\textbf{order}}.
\end{equation}
The Lagrange multiplier $k^{\alpha\beta\lambda}$ that imposes the metricity condition corresponds to the canonical conjugate momentum of the metric in CCGG. (See Ref. \cite{Struckmeier:2017vkf}. For the correspondence between the first and the second-order formalism in the torsion-free case see also \cite{Benisty:2018efx}.)

In previous CCGG formulations, a special Ansatz for the Hamiltonian structure 
up to second order in $\tilde{q}\indices{_{\lambda}^{\alpha\beta\mu}}$ 
was considered that is compatible with the Schwarzschild metric. Here we take all possible invariants of the momentum fields, i.e. Riemann tensors in the Lagrangian picture, into account, exploring the advantages of the second-order formalism. We prove that, within the canonical transformation framework,  the presence of the quadratic invariant in the Hamiltonian is necessary, and the Hamiltonian of the CCGG theory \cite{Struckmeier:2017vkf} is the minimal extension of the Einstein-Hilbert Ansatz. 

The manuscript is organized as follows: The first chapter reviews the basic principles of the Covariant Canonical Gauge theory of Gravity and calculates the equations of motion in the first formalism without assuming torsion. The second chapter states that if the $\mathcal{H}_{\textbf{dyn}}$ does not depend on the metric conjugate momenta, the field equations of motion correspond to the field equations in the second order formalism, because of the correspondence between the formulation theorem (Eq. (\ref{t})) \cite{Benisty:2018fgu}. The energy momentum conservation is also guarantied. Under those conditions, the third chapter implements the generalized combination of the $q^{\alpha\beta\gamma}_{\delta}$ tensors for the $\mathcal{H}_{\textbf{dyn}}$, which yields to the quadratic curvatures terms after Legendre transformation. The fourth chapter contains conformal invariant extensions into the CCGG. The last chapter summarizes the results of the paper and discuss possible future work.    

%
%
%
%
%
%
%
%
%
%
%
%
%
%
%
%
%
%
%
%
%
%
%
%

\section{Covariant Hamiltonian formulation}
\subsection{The gauge field}
The starting point of the CCGG framework is a globally Lorentz invariant Lagrangian and the corresponding action integral for classical matter fields.  The Legendre transform of the Lagrangian, the Hamiltonian,  depends on the fields and their conjugates. The conjugate momentum components of the fields are the duals of the complete set of the derivatives of the field in the Hamiltonian, including dynamic metric. 

The key element of the canonical transformation  framework, enforcing invariance  of a system's action integral with respect to some local transformation (Lie) group,  are the so called generating functions fixing the transformation law (covariance) of the matter fields. Form invariance of the Hamiltonian, though, can only be achieved by introducing compensating gauge fields, in analogy  to the electromagnetic field enabling the local phase invariance pertinent to local U(1) symmetry. In consequence, the original matter Hamiltonian is modified, in essence by replacing the partial derivatives by covariant derivatives. This methodology has been applied previously to the SU(N) group \cite{Struckmeier:2016kzt} and shown to reproduce the known Yang Mills theories of the electroweak and strong interactions. (Obviously, in order to apply that  framework operating in the Hamiltonian picture we must request the very existence of the Hamiltonian, i.e. the regularity of the Lagrangian). 
In CCGG, the symmetry group in question is the diffeomorphism group representing the General Principle of Relativity. Diffeomorphism are general coordinate transformations ($x^{\mu} \rightarrow X^{\mu}$), under which the fields transform as tensors, e.g. 
\begin{equation}\label{eq:trans-met}
G_{\mu\nu}(X)=g_{\alpha\beta}(x)\pfrac{x^{\alpha}}{X^{\mu}}\pfrac{x^{\beta}}{X^{\nu}}
\end{equation}
holds for the metric tensor. (The transformed quantities in the coordinate system $X$ are denoted by capital letters.) 
The gauge field  is the affine connection $\gamma\indices{^{\eta}_{\alpha\xi}}$ with the transformation law set up to ensure the invariance of the Hamiltonian:
\begin{equation}\label{eq:trans-conn-coeff}
\Gamma\indices{^{\kappa}_{\alpha\beta}}(X)=
\gamma\indices{^{\xi}_{\eta\tau}}(x)\pfrac{x^{\eta}}{X^{\alpha}}
\pfrac{x^{\tau}}{X^{\beta}}\pfrac{X^{\kappa}}{x^{\xi}}+
\ppfrac{x^{\xi}}{X^{\alpha}}{X^{\beta}}\pfrac{X^{\kappa}}{x^{\xi}}.
\end{equation}

The specific generating function implementing field transformations like \eqref{eq:trans-conn-coeff} - \eqref{eq:trans-met} under general coordinate transformations (diffeomorphism) have been introduced in Ref \cite{Struckmeier:2017vkf}. The resulting action becomes then a world scalar with partial derivatives converted to covariant derivatives (For fermions, the process and the resulting amendments are a bit more complex, though) and  partial derivatives of the connection replaced by the Riemann-Christoffel curvature tensor. The gravitational portion of the action then becomes: 
\begin{equation}
S_G=\int_{R}\left(\tilde{k}^{\,\alpha\lambda\beta}\,g_{\alpha\lambda;\beta}-\onehalf\tilde{q}\indices{_{\eta}^{\alpha\xi\beta}}R\indices{^{\eta}_{\alpha\xi\beta}} -\HCdtD \right)d^{4}x.
\label{eq:action-integral4}
\end{equation}
The ``tilde'' sign denotes a tensor density, where the tensor is multiplied by $\sqrt{-g}$.  The dynamical Hamiltonian $\HCdtD$, which is supposed to describe the dynamics of the free gravitational field, is not determined by the gauge process. It is to be built from a combination of the metric conjugate momenta $\tilde{k}^{\alpha\beta\mu}$, the connection conjugate momenta $\tilde{q}\indices{^{\eta}_{\alpha\xi\beta}}$, and the metric $g_{\alpha\beta}$ based on physical insights.  

Here we will work on the assumption that the connection is symmetric under the permutation of its lower indices:
\begin{equation}
\gamma\indices{^{\kappa}_{\alpha\beta}} = \gamma\indices{^{\kappa}_{\beta\alpha}}.
\end{equation}
This restriction was omitted in \cite{Struckmeier:2017vkf, Vasak:2018gqn} which allows torsion to be present. It was shown (in Ref. \cite{Benisty:2018efx}) that with metric compatibility a symmetric connection ensures covariant conservation of the matter energy-momentum tensor. That restriction is possible as the transformation rule (\ref{eq:trans-conn-coeff}) required for the gauge field is satisfied with a symmetric connection. 
\begin{figure*}[t]
 	\centering
    \includegraphics[width=0.9\textwidth]{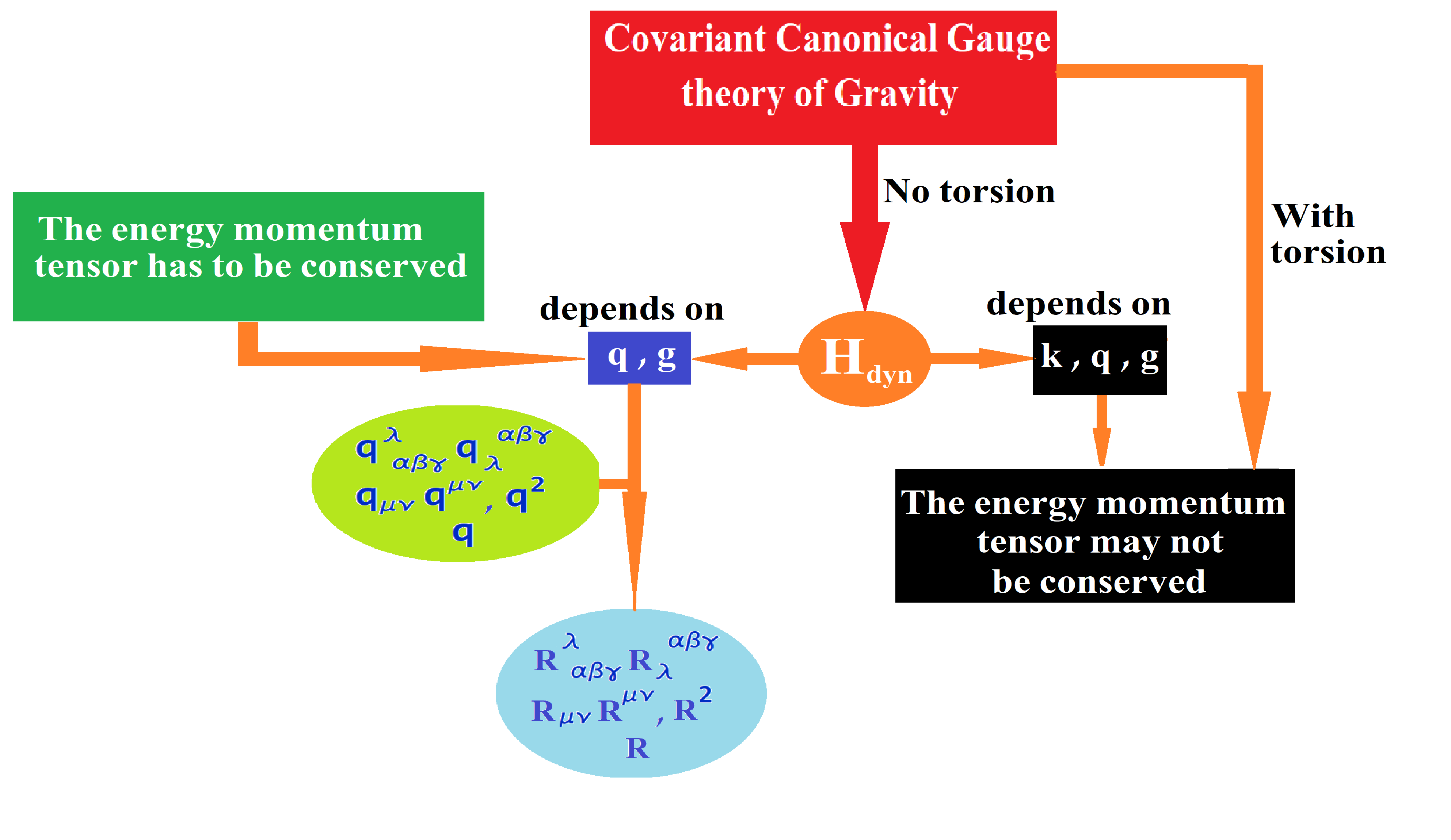}
\caption{Scheme of the connection between the dependence of the dynamical Hamiltonian on the momentum fields and the resulting theories of gravity. The energy momentum conservation for each case is also indicated.} 
\label{fig}
\end{figure*}
%

\subsection{The equations of motions}
The diffeomorphism invariant action integral can then, after adding the matter portion in Lagrangian form, be written as 
\begin{equation}\label{action}
\mathcal{S} = \int \left[\tilde{k}^{\alpha\beta\lambda} g_{\alpha\beta;\lambda} - \onehalf \tilde{q}\indices{_{\lambda}^{\alpha\beta\xi}}R\indices{^{\lambda}_{\alpha\beta\xi}}-\HCdtD (\tilde{q},\tilde{k},g) + \mathcal{\tilde{L}}_m \right]d^4x
\end{equation}
The canonical equations of motion are derived by variation. 

The first variation with respect to the momentum field conjugate to the metric tensor  yields
\begin{equation}\label{k}
g_{\alpha\beta;\lambda} = \pfrac{\HCdtD}{\tilde{k}^{\alpha\beta\lambda}}. \end{equation}
Metric compatibility is ensured if the r.h.s. of the above equation vanishes. This is the case with $\HCdtD$ not depending on $\tilde{k}$, i.e. if $\tilde{k}$ is a cyclic variable. This case will be discussed in the next section.

The variation with respect to the symmetric connection yields
\begin{equation}\label{varCon}
-\left(\tilde{k}^{\alpha\mu\nu}+\tilde{k}^{\alpha\nu\mu}\right) g_{\alpha\rho} = \onehalf\nabla_{\beta} \left( \tilde{q}\indices{_{\rho}^{\mu\beta\nu}}+ \tilde{q}\indices{_{\rho}^{\nu\beta\mu}} \right),
\end{equation}
which is a relation between the momentum of the metric and the connection.

In order to isolate the tensor $k^{\mu\nu\lambda}$ one can use the following procedure: First, we multiply by the metric $g^{\rho\sigma}$ and sum over the index $\sigma$:
\begin{equation}\label{varCon1}
-\tilde{k}^{\sigma\mu\nu}-\tilde{k}^{\sigma\nu\mu}= \onehalf\nabla_{\alpha} \left( \tilde{q}^{\,\,\sigma\mu\alpha\nu}+ \tilde{q}^{\,\,\sigma\nu\alpha\mu} \right).
\end{equation}
Switching the indices $\sigma \leftrightarrow \nu$:
\begin{equation}\label{varCon2}
-\tilde{k}^{\nu\mu\sigma}-\tilde{k}^{\nu\sigma\mu}= \onehalf\nabla_{\alpha} \left( \tilde{q}^{\nu\mu\alpha\sigma}+ \tilde{q}^{\nu\sigma\alpha\mu} \right),
\end{equation}
and the indices $\mu\leftrightarrow \nu$:
\begin{equation}\label{varCon3}
-\tilde{k}^{\mu\nu\sigma}-\tilde{k}^{\mu\sigma\nu}= \onehalf\nabla_{\alpha} \left( \tilde{q}^{\mu\nu\alpha\sigma}+ \tilde{q}^{\mu\sigma\alpha\nu} \right).
\end{equation}
Combining the equations  (\ref{varCon1}) $+$ (\ref{varCon2}) $-$ (\ref{varCon3}) we can isolate the tensor
\begin{equation}\label{varCovf}
\tilde{k}^{\sigma\nu\mu} = -\onehalf \nabla_\alpha \left(\tilde{q}^{\sigma\mu\alpha\nu}+\tilde{q}^{\nu\mu\alpha\sigma}\right).
\end{equation}

The third variation is with respect to the conjugate momentum $\tilde{q}\indices{_{\sigma}^{\mu\nu\rho}}$ of the connection yielding
\begin{equation}
\pfrac{\HCdtD}{\tilde{q}\indices{_{\sigma}^{\mu\nu\rho}}} = -\onehalf R\indices{^{\sigma}_{\mu\nu\rho}}.
\end{equation}
Obviously if $\HCdtD$ does not depend on $\tilde{q}$, the Riemann tensor will be zero, as in Teleparallel gravity \cite{Bahamonde:2015zma}. The last variation is with respect to the metric: 
\begin{equation} \label{varmetric}
\begin{split}
T^{\mu\nu} =g^{\mu\nu}\left(-k^{\alpha\beta\gamma} g_{\alpha\beta;\gamma} + \onehalf q_{\lambda}^{\alpha\beta\gamma}R\indices{^{\lambda}_{\alpha\beta\gamma}}\right)\\
+2k^{\mu\nu\gamma}_{;\gamma} - \frac{2}{\sqrt{-g}}\pfrac{\HCdtD}{g_{\mu\nu}}.
\end{split}
\end{equation}
Here 
\begin{equation}
T^{\mu\nu} =: \frac{2}{\sqrt{-g}}\pfrac{\tilde{\mathcal{L}}_m}{g_{\mu\nu}}
\end{equation}
is the metric energy-momentum (stress) tensor of matter in balance with the metric  energy-momentum (strain) tensor of matter. Using Eq. (\ref{varCovf}) we can replace $k^{\alpha\beta\gamma}$ and its derivative:
\begin{equation}\label{emtl}
\begin{split}
T^{\mu\nu} = g^{\mu\nu}\left[\onehalf \nabla_\sigma \left(\tilde{q}^{\alpha\gamma\sigma\beta}+\tilde{q}^{\beta\gamma\sigma\alpha}\right) g_{\alpha\beta;\gamma} + \onehalf q_{\lambda}^{\alpha\beta\gamma}R\indices{^{\lambda}_{\alpha\beta\gamma}}\right] \\- \frac{2}{\sqrt{-g}}\pfrac{\HCdtD}{g_{\mu\nu}} 
-\nabla_{\gamma}\nabla_{\alpha}\left(q^{\mu\gamma\nu\alpha} + q^{\nu\gamma\mu\alpha}\right).
\end{split}
\end{equation}
From the correspondence between the formulations (Eq. (\ref{t})), even the action is derived in Palatini formalism, the field equations equal to the field equations in the metric formalism.  
\section{Correspondence between the $1^{st}$ and the $2^{nd}$ order formalism}
\subsection{$\HCdtD(q,g)$ with metric compatibility}
A particular case of $\HCdtD(k,q,g)$ is when the metric conjugate momentum $k^{\alpha\beta\gamma}$ does not occurs in $\HCdtD(q,g)$. From the variation with respect to the metric conjugate momentum $k^{\alpha\beta\gamma}$, Eq. (\ref{k}), we get the metric compatibility condition. In conjunction with neglecting torsion  the connection is Levi-Civita, i.e. equal to the Christoffel symbol:
\begin{equation}\label{met}
g_{\alpha\beta;\gamma} = 0 \quad \Rightarrow \quad \gamma\indices{^{\rho}_{\mu\nu}} = \left\{ \genfrac{}{}{0pt}{}{\rho}{\mu \nu} \right\}.
\end{equation}
In Ref. \cite{Benisty:2018fgu} it was proven that for any general action which starts in the first-order formalism with the term $k^{\alpha\beta\gamma}  g_{\alpha\beta;\gamma}$ added as a Lagrange multiplier, the strain tensor is identical to that derived via the second-order formalism. The main reason for that correspondence is the variations of the term $k^{\alpha\beta\gamma}  g_{\alpha\beta;\gamma}$. While, as mentioned above, the variation with respect to $k^{\alpha\beta\gamma}$ ensures metricity, the variation with respect to the connection yields
\begin{equation}\label{g}
\frac{\partial}{\partial \gamma\indices{^{\rho}_{\mu\nu}}}  k^{\alpha\beta\lambda}  g_{\alpha\beta;\lambda} = - k^{\alpha\mu\nu}g_{\rho\alpha} - k^{\alpha\nu\mu}g_{\rho\alpha}
\end{equation}
with a symmetrization between the components $\mu$ and $\nu$. Moreover, the variation with respect to the metric yields
%
\begin{equation}\label{eomm}
\frac{\partial}{\partial g_{\mu\nu}}  k^{\alpha\beta\lambda}  g_{\alpha\beta;\lambda} = -k\indices{^{\mu\nu\lambda}_{;\lambda}}.
\end{equation}
$k\indices{^{\mu\nu\lambda}_{;\lambda}}$ contributes to the field equation \eqref{varmetric}, hence, the first order field equations turns out to be equivalent to the field equations in the second order formalism. Indeed, isolating the tensor $k^{\mu\nu\lambda}$ and inserting it back into Eq.~(\ref{eomm}) leads to the relation: 
%
\begin{equation}\label{connection}
\pfrac{\mathcal{L}(\kappa)}{g_{\sigma\nu}} = \onehalf\nabla_{\mu}\left(g^{\rho\sigma}\pfrac{ \mathcal{L}(\kappa)}{\gamma\indices{^{\rho}_{\mu\nu}}}+g^{\rho\nu}\pfrac{\mathcal{L}(\kappa)}{\gamma\indices{^{\rho}_{\mu\sigma}}}-g^{\rho\mu}\pfrac{\mathcal{L}(\kappa)}{\gamma\indices{^{\rho}_{\nu\sigma}}}\right),
\end{equation}
where $\mathcal{L}(\kappa) = k^{\alpha\beta\gamma}  g_{\alpha\beta;\gamma}$.
%
%
The terms on the right-hand side represent the additional terms that appears in the second-order formalism. The strain energy-momentum tensor (\ref{emtl}) then gets the additional contribution,
\begin{equation}
- \nabla_{\gamma}\nabla_{\alpha}\left(q^{\mu\gamma\nu\alpha} + q^{\nu\gamma\mu\alpha}\right). 
\end{equation}
This is exactly the additional contribution for the field equation of motion (\ref{emtl}) which yields the missing terms that arenot present in the first order formalism.
\subsection{Energy momentum conservation}
The strain energy-momentum tensor may not be covariantly conserved in the first-order formalism, in contrast to the second-order formalism, where the energy-momentum tensor must be conserved \cite{Borunda:2008kf}. An important fundamental link between the dependence of the $\HCdtD$ with the metric conjugate momentum $\tilde{k}^{\alpha\beta\gamma}$ and the conservation of the stress energy tensor is obtained, through the theorem of the correspondence between the first and second-order formalism. In the particular case, if $\HCdtD$ does not depend on the metric conjugate momentum $\tilde{k}^{\alpha\beta\gamma}$:
%
%
\begin{equation} \label{SA}
S_G^{metric}=\int_{R}\left(\tilde{k}^{\,\alpha\lambda\beta}\,g_{\alpha\lambda;\beta}-\onehalf\tilde{q}\indices{_{\eta}^{\alpha\xi\beta}}R\indices{^{\eta}_{\alpha\xi\beta}}-\HCdtD(\tilde{q},g)\right)d^{4}x.
\end{equation}
A variation with respect to the metric conjugate momentum $\tilde{k}^{\alpha\beta\gamma}$ yields the metric compatibility condition. According to the theorem (\ref{t}), the gravitational stress energy-momentum tensor is the same as the strain tensor in the second-order formalism, which ensures the covariant conservation of this strain tensor:
\begin{equation}\label{conserved}
\nabla_{\mu} G^{\mu\nu} = 0, \qquad  G^{\mu\nu} =: -\frac{2}{\sqrt{-g}}\frac{\delta S_G^{metric}}{\delta g^{\mu\nu}}.
\end{equation}
In the generic case $\HCdtD$ does depend on the metric conjugate momentum $\tilde{k}^{\alpha\beta\mu}$:
\begin{equation}
S_G=\int_{R}\left(\tilde{k}^{\,\alpha\lambda\beta}\,g_{\alpha\lambda;\beta}-\onehalf\tilde{q}\indices{_{\eta}^{\alpha\xi\beta}}R\indices{^{\eta}_{\alpha\xi\beta}}-\HCdtD (\tilde{q},\tilde{k},g)\right)d^{4}x.
\end{equation}
Then the variation with respect to the metric conjugate momentum $\tilde{k}^{\alpha\beta\mu}$ breaks the metric compatibility condition, and the gravitational strain energy tensor may not be covariantly conserved. Of course, this applies also if the assumption of zero torsion is dropped. This basic framework is not a special feature only for CCGG but leads to a fundamental correlation for many options for $\HCdtD$. Fig.~(\ref{fig}) summarizes the links between the formulation of the theory and the covariant conservation of the strain energy-momentum tensor.
%

\section{Generic quadratic invariants in $q\indices{_{\eta}^{\alpha\beta\mu}}$}

\subsection{Complete combination of the $q$ tensors}
The term  $R\indices{_{\delta}^{\alpha\beta\lambda}}q\indices{^{\delta}_{\alpha\beta\lambda}}$ in the action \eqref{eq:action-integral4} contributes only if the Riemann tensor and the $q\indices{_{\delta}^{\alpha\beta\lambda}}$ tensor have the same symmetries and anti-symmetries. Hence we may build similar contractions and combinations of the conjugate momenta of the connection, as are possible with the Riemann tensor. For example $q^{\alpha\beta}$ in analogy to the Ricci tensor:
\begin{equation}
R^{\mu\nu} = R\indices{_{\lambda}^{\mu\lambda\nu}}, \quad q^{\mu\nu} = q\indices{_{\lambda}^{\mu\lambda\nu}},
\end{equation}
or the $q$ scalar in analogy to the Ricci scalar:
\begin{equation}
R = R^{\mu\nu}g_{\mu\nu}, \quad  q = q^{\mu\nu}g_{\mu\nu}.
\end{equation}
Those fundamental definitions will be used to built the general form of the free gravity Hamiltonian $\HCdtD$ . 
\subsection{The general form}
In order to see the complete implications for this formulation, our starting point is a dynamical Hamiltonian with the connection conjugate momentum $q\indices{_{\delta}^{\alpha\beta\lambda}}$ up to the second order, but without a dependence on the metric conjugate momentum $k\indices{^{\alpha\beta\lambda}}$:
\begin{equation}\label{DH}
\HCdD = \mathcal{H}_{0} + \mathcal{H}_{1} + \mathcal{H}_{21} + \mathcal{H}_{22} + \mathcal{H}_{23}
\end{equation}
\begin{subequations}
\begin{equation}
\mathcal{H}_0 = g_0
\end{equation}
\begin{equation}
\mathcal{H}_{1} = - \frac{g_1}{2} q 
\end{equation}
\begin{equation}
\mathcal{H}_{21} = -\frac{1}{4} g_{21}\, q^2
\end{equation}
\begin{equation}
\mathcal{H}_{22} = -\frac{1}{4} g_{22} \, q^{\mu\nu}q_{\mu\nu}
\end{equation}
\begin{equation}
\mathcal{H}_{23} = -\frac{1}{4} g_{23}\, q^{\alpha\beta\gamma\delta}q_{\alpha\beta\gamma\delta} 
\end{equation}
\end{subequations}
That Hamiltonian contains all possible (up to a sign) combinations of $q$ tensors. As $\HCdD$ does not depend on  $k^{\alpha\beta\gamma}$, the existence of the term $k^{\alpha\beta\lambda}g_{\alpha\beta;\lambda}$ ensures, as discussed above, a correspondence between the first and the second-order formalism. In consequence, the strain tensor is covariantly conserved, i.e. Eq. (\ref{conserved}) holds.
\subsection{The $q\indices{_{\sigma}^{\mu\nu\rho}}$ tensor}
In order to derive the relation between the momentum field $q\indices{_{\sigma}^{\mu\nu\rho}}$ and the Riemann curvature, we carry out the corresponding variation, and obtain:
\begin{equation}\label{ten}
R\indices{_{\sigma}^{\mu\nu\rho}} = g_{23} q\indices{_{\sigma}^{\mu\nu\rho}} + \delta^{\nu}_{\sigma} \left(g_{22} q^{\mu\rho} + g^{\mu\rho} (g_{21} q + g_1) \right).
\end{equation}
To isolate the $q$ tensor, we take the trace of this equation by  contracting with $\delta^{\sigma}_{\nu}$:
\begin{equation}\label{symten}
R^{\mu\rho} = (g_{23}+4g_{22}) q^{\mu\rho} + 4g^{\mu\rho} (g_{21} q + g_{1}). 
\end{equation}
Taking trace of this equation by contracting with $g_{\mu\rho}$ gives:
\begin{equation}\label{sca}
R= (g_{23} + 4g_{22} + 16 g_{21}) q + 16 g_{1}.
\end{equation}
Solving for $q$, 
\begin{equation}\label{sca2}
q = \frac{R - 16 g_{1}}{g_{23} + 4g_{22} + 16 g_{21}},
\end{equation}
and plugging $q$ into Eq. (\ref{symten}) allows to extract the $q^{\mu\nu}$ tensor in terms of the metric and the Ricci tensor and scalar: 
\begin{equation}\label{qmunu}
\begin{split}
q^{\mu\nu} = \frac{R^{\mu\nu}}{4 g_{22} + g_{23}} - \frac{4 g_1}{16 g_{21} + 4 g_{22} + g_{23}} g^{\mu\nu} \\ -\frac{4 g_{21}}{(4 g_{22} + g_{23})(16 g_{21} + 4 g_{22} + g_{23})}R g^{\mu\nu} .
\end{split}
\end{equation}
In the last step we insert the relations for $q$ and $q^{\mu\nu}$ into Eq. (\ref{ten}), and reconstruct the momentum tensor from the Riemann curvature tensor:
\begin{equation}\label{q}
\begin{split}
\boxed{q\indices{_{\sigma}^{\mu\nu\rho}} = \alpha_1 R\indices{_{\sigma}^{\mu\nu\rho}} + \delta^{\nu}_{\sigma}\left(\alpha_2 R^{\mu\sigma} + \alpha_3 R g^{\mu\sigma} + \alpha_4 g^{\mu\sigma} \right)}.
\end{split}
\end{equation}
The new coupling constant are combinations of the original ones:
\begin{subequations}\label{alpha}
\begin{equation}
\alpha_1 = \frac{1}{g_{23}}
\end{equation}
\begin{equation}
\alpha_2 = -\frac{g_{22}}{g_{23} (4 g_{22}+g_{23})}
\end{equation}
\begin{equation}
\alpha_3 = -\frac{g_{21}}{(4 g_{22}+g_{23}) (16 g_{21}+4 g_{22}+g_{23})}
\end{equation}
\begin{equation}
\alpha_4 = -\frac{g_1}{16 g_{21}+4 g_{22}+g_{23}}.
\end{equation}
\end{subequations}
Now the Ricci tensor and scalar read: 
\begin{equation}
q^{\mu\nu} = \left(\alpha_1 + 4 \alpha_2 \right) R^{\mu\nu}+ 4\left(\alpha_3 g^{\mu\nu}R + \alpha_4 g^{\mu\nu}  \right),
\end{equation}
\begin{equation}
q = \left(\alpha_1 + 4\alpha_2 + 16 \alpha_3\right)R + 16\alpha_4.
\end{equation}
Eq. (\ref{alpha}) shows that the reconstruction of the momentum tensor is possible only if $g_{23} \ne 0$, i.e. the success of the canonical formulation depends on the existence of the $q^{\alpha\beta\gamma\delta}q_{\alpha\beta\gamma\delta}$ term which, not surprisingly, also ensures the existence of the Legendre transform. This proves that the choice of $\HCdtD$ in CCGG \cite{Struckmeier:2017vkf} is the necessary but minimal extension of the Einstein-Hilbert theory.

\subsection{The quadratic Lagrangian}
Because of the variation with respect to $q\indices{_{\sigma}^{\mu\nu\rho}}$ gives the solution for $q\indices{_{\sigma}^{\mu\nu\rho}}$, and since $q\indices{_{\sigma}^{\mu\nu\rho}}$ is a field with zero canonical momentum, which does not contain derivatives, we can replace the solution for $q^{\alpha\beta\gamma}_\delta$ into the action. The Lagrangian is obtained by carrying out the Legendre transform given by the integrand in the action integral (\ref{SA}):
\begin{equation*}
\boxed{-\mathcal{L}= \frac{\alpha_1}{4} R\indices{_{\lambda}^{\alpha\beta\gamma}}R\indices{^{\lambda}_{\alpha\beta\gamma}} + \frac{\alpha_2}{4} R^{\mu\sigma} R_{\mu\sigma} + \frac{\alpha_3}{4} R^2 + \frac{\alpha_4}{2}R + \Lambda}.
\end{equation*}
The emerging "cosmological constant" $\Lambda$ is defined by 
\begin{equation}
\Lambda =  -4g_1 \alpha_4 + g_0 = -\frac{4g_1^2}{16 g_{21} + 4 g_{22}+ g_{23}} + g_0.
\end{equation}
This formula for the cosmological constant facilitates an option for resolving the cosmological constant problem as suggested earlier in Ref. \cite{Vasak:2018gqn} for a dynamical Hamiltonian with
\begin{equation}
g_{21} = g_{22} = g_0 = 0.
\end{equation}
Then one considers a very large $|g_{23}|$ and $g_{23}<0$. For that case, the cosmological constant $\Lambda$ is highly suppressed. This is similar to the "cosmological see saw mechanism" to obtain a low value of the effective cosmological constant \cite{Guendelman:1999qt}. In order to get the small cosmological constant we could require $|16 g_{21} + 4 g_{22}+ g_{23}|$ being very large, and $16 g_{21} + 4 g_{22}+ g_{23}<0$. This applies also for $g_0 = 0$. 

The potential of the quadratic term to transfer energy from gravity to matter and vice versa and generate inflation \cite{Starobinsky:1980te} has been shown in Ref. \cite{Benisty:2018ywz}.
\subsection{Special cases}
There are two particularly interesting special cases for the coupling constant. For the Gauss-Bonnet combination \cite{Lovelock:1971yv} of the quadratic terms, the original $g$ constants must fulfill the relation
\begin{equation}
g_{21} = \xi, \quad g_{22} = -4\xi, \quad g_{23} = 15\xi,
\end{equation}
where $\xi$ is some free constant. For this combination we get the condition for the $\alpha$ couplings:
\begin{equation}
\alpha_1 = -\frac{\alpha_2}{4} = \alpha_3.
\end{equation}
For the Conformal Gravity \cite{Mannheim:2011ds}-\cite{Mannheim:2005bfa} based on the quadratic Weyl tensor, the original $g$ constants are chosen to be:
\begin{equation}
\begin{split}
g_{21} = \xi, \quad g_{22}= 10 \xi ,\quad g_{23}=-35 \xi 
\\ \quad g_{0} = g_1 = 0
\end{split}
\end{equation}
where $\xi$ is some free constant. After the Legendre transform, the final coupling constants are:  
\begin{equation}
\alpha_1 = -\frac{\alpha_2}{2} = 3\alpha_3, \quad \Lambda = 0
\end{equation}
which gives eventually the quadratic Weyl tensor in the effective action:
\begin{equation}
\mathcal{L} = \frac{\xi}{4} C_{\mu\nu\alpha\beta}C^{\mu\nu\alpha\beta},
\end{equation}
that is the familiar conformal invariant action. 

\section{A conformal invariant extension}
%
%
An extension which breaks the metric compatibility condition and respects conformal invariance \cite{weyl}-\cite{Romero:2012hs} can be developed based on the above considerations  by introducing a vector field $A_\mu$ into the metricity constraint:
\begin{equation}
k^{\alpha\beta\gamma}  g_{\alpha\beta;\gamma} \Rightarrow k^{\alpha\beta\gamma}  (g_{\alpha\beta;\gamma}- e g_{\alpha\beta}A_{\gamma}).
\end{equation}The corresponding generalized action then reads:
\begin{equation}
\begin{split}
\mathcal{S} = \int [\tilde{k}^{\alpha\beta\gamma}  (g_{\alpha\beta;\gamma}- e g_{\alpha\beta}A_{\gamma}) \\ -\onehalf \tilde{q}\indices{_{\lambda}^{\alpha\beta\xi}}R\indices{^{\lambda}_{\alpha\beta\xi}}-\HCdtD (\tilde{q},\tilde{k},g) + \mathcal{\tilde{L}}_m ] d^4x,
\end{split}
\end{equation}
where $e$ is the "conformal charge" of the conformal gauge field~$A_\mu$. Assuming that the connection will be covariant under conformal transformation, the symmetries give:
%
%
\begin{equation}
\Gamma^{\lambda}_{\alpha\beta} \rightarrow \Gamma^{\lambda}_{\alpha\beta} , \quad g_{\mu\nu} \rightarrow \Omega(x^\mu)^2 g_{\mu\nu} , \quad k_{\alpha\beta\gamma} \rightarrow k_{\alpha\beta\gamma}
\end{equation}
\begin{equation*}
A_\mu \rightarrow A_\mu + \frac{2}{e} \partial_\mu \log \Omega(x^\mu) , \quad q\indices{_{\lambda}^{\alpha\beta\gamma}} \rightarrow  \Omega(x^\mu)^{-4} q\indices{_{\lambda}^{\alpha\beta\gamma}}
\end{equation*}
that the metric conjugate momenta $k_{\alpha\beta\gamma}$ with lower indices does not transform. From the variation of $k_{\alpha\beta\gamma}$ the condition of Weyl's non-metricity is obtained from the action:
\begin{equation}\label{nmc}
\nabla_\gamma g_{\alpha\beta} = e A_\gamma g_{\alpha\beta}
\end{equation} 
which leads to the solution for the connection:
\begin{equation}
\Gamma^{\rho}_{\mu\nu} = \left\{ \genfrac{}{}{0pt}{}{\rho}{\mu \nu} \right\} - \frac{e}{2} g^{\rho\lambda} (g_{\lambda\mu}A_{\nu}+g_{\lambda\nu}A_{\mu}-g_{\mu\nu}A_{\lambda})
\end{equation}
For conformal invariant dynamical Hamiltonian:
\begin{equation}
\mathcal{H}_\textbf{Dyn} \rightarrow \mathcal{H}_\textbf{Dyn}
\end{equation}
only quadratic terms of the $q\indices{_{\sigma}^{\mu\nu\rho}}$ tensors are kept, with the coupling constants of the Hamiltonian Eq. (\ref{DH}):
\begin{equation}
g_{0} = g_{1} = 0.
\end{equation}
In addition, a kinetic term based on the gauge field $A_\mu$ could be added:
\begin{equation}
\mathcal{L}^{(\textrm{Kin})}=-\frac{1}{4} F_{\mu\nu}F^{\mu\nu}
\end{equation}.
Because of the new contributions, the complete field equation is Weyl invariant. This subject could be study in details in the future.

In order to have a linear term in curvature which respects the conformal invariance symmetry, one can use a modified measure, which is independent of the metric in addition to the action that was discussed \cite{Guendelman:1996jr}:
\begin{equation}
\mathcal{S} = \int d^4 x \, \Phi R
\end{equation}
A simple construction of this modified measure is from four scalar fields $ \varphi_{a} $, where $ a=1,2,3,4 $. 
\begin{equation}
\Phi=\frac{1}{4!}\varepsilon^{\alpha\beta\gamma\delta}\varepsilon_{abcd}\partial_{\alpha}\varphi^{(a)}\partial_{\beta}\varphi^{(b)}\partial_{\gamma}\varphi^{(c)}\partial_{\delta}\varphi^{(d)},
\end{equation}  
with the conformal symmetry being realized by the transformation:
\begin{equation}
\varphi_a'\rightarrow\varphi_a'(\varphi), \quad \Phi' \rightarrow \Phi \, \Omega(x)^2.
\end{equation}
where the Jacobian of the transformation being correlated with $\Omega$ through the relation $J = \Omega(x)^2$.
This is one option for breaking the metricity condition, and respecting the conformal invariance, using the same Lagrange multiplier.

\section{Conclusions}
The Covariant Canonical Gauge theory of Gravity (CCGG) is a classical covariant field theory in the Hamiltonian picture using the framework of canonical transformations to implement local invariance with respect to the diffeomorphism group.  Its fundamental ingredients  are the metric $g_{\mu\nu}$, the affine connection $\gamma\indices{^{\lambda}_{\mu\nu}}$ and their conjugate momenta, the $k^{\mu\nu\sigma}$ and the $q\indices{_{\eta}^{\alpha\xi\beta}}$ tensors. The metric is a dynamical field and the connection is the (independent) gauge field.  The final covariant Hamiltonian of the gauge theory is taken to be in the second order of the $q\indices{_{\eta}^{\alpha\xi\beta}}$ tensor. This approach seems well suited for the quantization of quadratic curvature theories, because it reduces the derivatives in the action. We hope to discuss the quantization of higher-derivative theories in the future, based on this formalism.

In this paper we have generalized torsion and metricity-compatible CCGG to account for all higher curvature invariants in the action up to second order in the $q\indices{_{\eta}^{\alpha\xi\beta}}$ tensor components, including the contractions ($q^{\mu\nu}q_{\mu\nu}$ and $q^2$). This leads to all possible combinations of second-order curvature invariants in the Lagrangian. It turns out that the quadratic $q\indices{_{\eta}^{\alpha\xi\beta}}$ tensor invariant must be present in the Hamiltonian in order to make the theory consistent, otherwise $g_{23}=0$, and all of the $\alpha$'s coupling constants contain singularities (See Eq. (\ref{alpha})). 

CCGG can also be extended with more general dynamical Hamiltonian depending on the metric conjugate momenta $k^{\alpha\beta\gamma}$. In this case, the metric compatibility condition will be violated and a gate for new physics opened.  A simple example for such a theory with non-metricity coupled to a vector field is shown to maintain conformal invariance. These options and others could be investigated in the future.
\bigskip
\acknowledgments
David Benisty thanks to Mrs.\ Margarethe Puschmann and the Herbert Puschmann Stiftung for FIAS in the Verein der Freunde and Foerderer der Goethe University. David Vasak thanks the Carl-Wilhelm Fueck Stiftung for generous support through the Walter Greiner Gesellschaft zur Foerderung der physikalischen Grundlagenforschung Frankfurt. Eduardo Guendelman and David Benisty thank for the support of COST Action CA15117 "Cosmology and Astrophysics Network for Theoretical Advances and Training Action" (CANTATA) of the COST (European Cooperation in Science and Technology). Horst Stocker thanks to the WGG and the Goethe University for the support through the Judah Moshe Eisenberg Laureatus endowed professorship and thanks the BMBF (German Federal Ministry of Education and Research).

\end{document}